      [1999/12/01 v1.4c Il Nuovo Cimento]
\documentclass{cimento}

\usepackage{amssymb}
\usepackage{amsmath}
\usepackage{graphicx}
\usepackage{esint}

\newcommand{\ch}{\mathrm{ch}}

\newcommand{\beqn}{\begin{eqnarray}}
\newcommand{\eeqn}{\end{eqnarray}}
\newcommand{\eq}[1]{(\ref{#1})}
\newcommand{\bl}{{\biggl|}}

\newcommand{\cL}{{\cal L}}

\newcommand{\Z}{{\mathbb{Z}}}

\title{Zero-point fluctuations in rotation: Perpetuum mobile of the fourth kind without energy transfer\thanks{Talk given at the workshop ``Mathematical Structures in Quantum Systems and applications'',  08 -- 14 July, 2012, Benasque, Spain.}
}

\author{M. N. Chernodub\thanks{On leave from ITEP, Moscow, Russia.}}
\instlist{\inst CNRS, Laboratoire de Math\'ematiques et Physique Th\'eorique, 
Universit\'e Fran\c{c}ois-Rabelais Tours,\\ F\'ed\'eration Denis Poisson, Parc de Grandmont, 37200 Tours, France \\[1mm]
Department of Physics and Astronomy, University of Gent, Krijgslaan 281, S9, B-9000 Gent, Belgium}
\PACSes{
\PACSit{03.70.+k}{Quantized fields}
\PACSit{42.50.Lc}{Fluctuation phenomena, quantum optics}
\PACSit{42.50.Pq}{Cavity quantum electrodynamics}
}

\begin{document}

\maketitle

\begin{abstract}
We discuss a simple Casimir-type device for which the rotational energy reaches its global minimum when the device rotates about a certain axis rather than remains static. This unusual property is a direct consequence of the fact that the moment of inertia of zero-point vacuum fluctuations is a negative quantity (the rotational vacuum effect). Moreover, the device does not produce any work despite the fact that its equilibrium ground state corresponds to a permanent rotation. Counterintuitively, the device has no internally moving mechanical parts while its very existence is consistent with the laws of thermodynamics. We point out that such devices may possibly be constructed using carbon nanotubes. We call this ``zero-point-driven'' device as the perpetuum mobile of the fourth kind.
\end{abstract}

\clearpage

\section{Introduction}

\rightline{\it Perpetual motion: Motion\,  that\,  continues\,  indefinitely\,  without\,  any\, external\, } 
\rightline{\it source of energy; impossible in practice because of friction.} 
\vskip 1mm
\rightline{Webster's online dictionary~\cite{ref:Webster}}

\vskip 5mm

In the quoted Webster's definition, the moving object is assumed to be coupled to (or, in other words, interacting with) an external environment, otherwise the perpetual nature of the object's motion is a trivial and inevitable consequence of the energy conservation law. Due to the interaction with the environment, a moving (say, rotating) object will sooner or later release all its (rotational) kinetic energy to the surrounding media. The gradual loss of the rotational energy can be described by the presence of a friction force acting on the object from the environment. As a consequence, the object will ultimately stop its rotational motion coming to rest; hence the verdict "impossibility in practice" in the Webster's definition. 

In this talk, following Refs.~\cite{ref:I, ref:II}, we point out, that the perpetual motion may exist. The device, which we call ``Perpetuum mobile of the fourth kind''~\cite{ref:II}, is suggested to have the following physical properties:
\begin{itemize}

\item It is a large object made of the Avogadro-scale ($N \sim 10^{23}$) constituent particles.

\item It is a solid body which, counterintuitively, has no internally moving parts. 

\item It is made of a material, which possesses propagating (massless) excitations. 

\item The perpetual rotation of the device is literally driven by zero-point fluctuations of these excitations. The key observation is that the rotational zero-point energy of the excitations has an unusual (discontinuous and non-monotonic) dependence on the angular frequency.

\item The lowest energy state of the device corresponds to a uniform rotation of the device as a whole about a certain axis with a certain (optimal) angular frequency. In other words, a non-rotating device has a higher energy compared to its energy in the optimally rotating state.

\end{itemize}

Thus, we define the perpetuum mobile of the fourth kind as 
\begin{itemize}
\item[(i)] a solid macroscopic device for which 
\item[(ii)] the lowest-energy state corresponds to a uniform rotation 
\item[(iii)] driven by the zero--point fluctuations without any energy transfer. 
\end{itemize}

The energy properties of our perpetuum mobile are very counterintuitive, because in order to stop the rotation of the perpetuum mobile we should invest (not gain!) some amount of work in this process as the cessation of rotation increases (not decreases) the rotational kinetic energy of the device. In an elementary act of interaction of the rotating device with an external environment (say, with a molecule of a gas), the device
\begin{itemize}

\item[1.)] cannot loose energy and stop rotating because the rotating device is already residing in its lowest--possible energy state;

\item[2.)] can gain energy and stop rotating. 

\end{itemize}
Below, we sometimes call the perpetuum mobile of the fourth kind as ``the device''. 

There are certain systems/effects which exhibit a certain version of persistent motion and, thus, are philosophically similar to our device. 

The first well-known example is the persistent electric current in nanosized nonsuperconducting rings~\cite{ref:persistent}. The current is induced by a magnetic field which pierces the ring.

The second example appears naturally in various macroscopically coherent quantum systems such as superconductors and superfluids. The explicit examples include the persistent electric current in a superconducting ring and a persistent (metastable, but with huge lifetime) flow of superfluid Helium in a large closed ring-shaped pipe, respectively. 

The third example is a ``time crystal'' suggestion which was inspired by the persistency of the electric currents in superconducting rings~\cite{ref:Shapere:Wilczek:class,ref:Shapere:Wilczek:quant}. Mathematically, the ``time crystal'' is suggested to be an exotic classical, semiclassical or quantum--mechanical system of moving particles which is described by either singular Hamiltonians involving multiple powers of time derivatives~\cite{ref:Shapere:Wilczek:class}, or by Hamiltonians which involve time-nonlocal terms~\cite{ref:Shapere:Wilczek:quant}. This suggestion is both controversial~\cite{ref:comment} and inspiring. 

Recently, a very interesting ``space--time crystal'' system was suggested to be realised at cryogenic temperatures in a form a persistent ring-shaped current of cold ions~\cite{ref:CI}. Similarly to the persistent electric current in nanosized rings~\cite{ref:persistent}, the persistent current of cold ions is also induced by the magnetic field background.

Contrary to the ``time crystal'' approach, our alternative approach -- which we call ``the perpetuum mobile of the fourth kind''~\cite{ref:I,ref:II} -- is based on standard nonsingular class of local free (!) Hamiltonians of relatively simple solid state systems (which may, presumably, be constructed by using doped torus-shaped carbon nanotubes~\cite{ref:II}).

\section{Negative moment of inertia of zero-point fluctuations}

Let us assume for a moment that the perpetuum mobile of the fourth kind (described in the Introduction) does exist and a large macroscopic device rotates permanently in its ground (lowest) state without producing any work. The necessary condition of the permanent rotation is that the non-rotating state should have a higher energy compared to the optimally rotating state. What should be a driving mechanism behind this unusual energy behaviour? In the present Section we are sketching the answer to this question.


Let us consider a large macroscopic rigid body of an arbitrary shape which rotates about one of its a principal axis of inertia with the angular frequency $\Omega$. For the sake of simplicity let us consider a slow, nonrelativistic rotation. Then the classical kinetic energy of rotation is a quadratic function of the angular frequency:
\beqn
E_{\mathrm{cl}}(\Omega) = \frac{I_{\mathrm{cl}} \Omega^{2}}{2}\,,
\label{eq:E:class}
\eeqn
where $I^{\mathrm{cl}}$ is the corresponding classical principal moment of inertia. The energetically favourable state, $E^{\mathrm{cl}}_{\mathrm{rot}} = 0$, corresponds to the absence of rotation, $\Omega = 0$. 

Thus, the classical formula~\eq{eq:E:class} seems to imply that that the permanent rotation should not, in general, correspond to a lowest energy state. However, there is a hidden assumption which has led us to this mundane conclusion: We have implicitly assumed that the moment of inertia of the rigid body is a positive quantity:
\beqn
I_{\mathrm{cl}} > 0\,.
\label{eq:I:cl}
\eeqn

Can we question the validity of Eq.~\eq{eq:I:cl}? At the first sight it seems that the answer is ``no'' as the positiveness condition~\eq{eq:I:cl} is a very natural fact for any classical system. For example, consider a point mass $m$ rotating about a certain axis at the fixed distance $R$ (one can imagine that the point is attached to the axis by a massless rod). The rotational kinetic energy of the system is given by Eq.~\eq{eq:E:class}, where the moment of inertia is 
\beqn
I_{\mathrm{cl}} = m R^2\,.
\label{eq:I:cl:m}
\eeqn
The classical mass $m$ and the distance $R$ are positive quantities so that the moment of inertia~\eq{eq:I:cl:m} is, naturally, a positive quantity too~\eq{eq:I:cl}. Since any classical body can be considered as a superposition of point masses, the positivity condition~\eq{eq:I:cl} for all extended classical bodies is also an obvious fact.


However, the classical-like relation~\eq{eq:I:cl} is generally not true for quantum systems~\cite{ref:I}.

In Refs.~\cite{ref:I,ref:II} we claim that there may exist certain systems which have a negative moment of inertia for, at least, small angular frequencies. These systems contain, generally, two interacting (classical and quantum) subsystems. The classical subsystem is a rigid body which has a positive moment of inertia. The quantum subsystem is described by zero-point fluctuations of quantum excitations which have, in general, a negative moment of inertia (the latter is the essence of the ``rotational vacuum effect''~\cite{ref:I}). 

%
\begin{figure}[!thb]
\begin{center}
\begin{tabular}{rl}
\\[1mm]
\includegraphics[scale=0.30,clip=false]{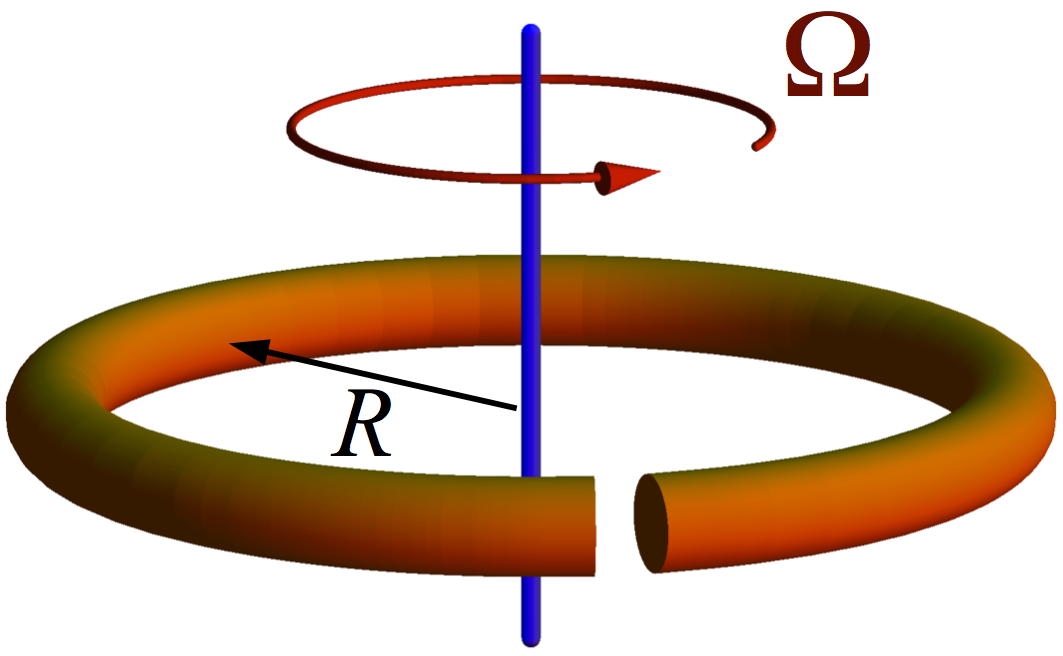} &  \\[-37mm]
 & \includegraphics[scale=0.18,clip=false]{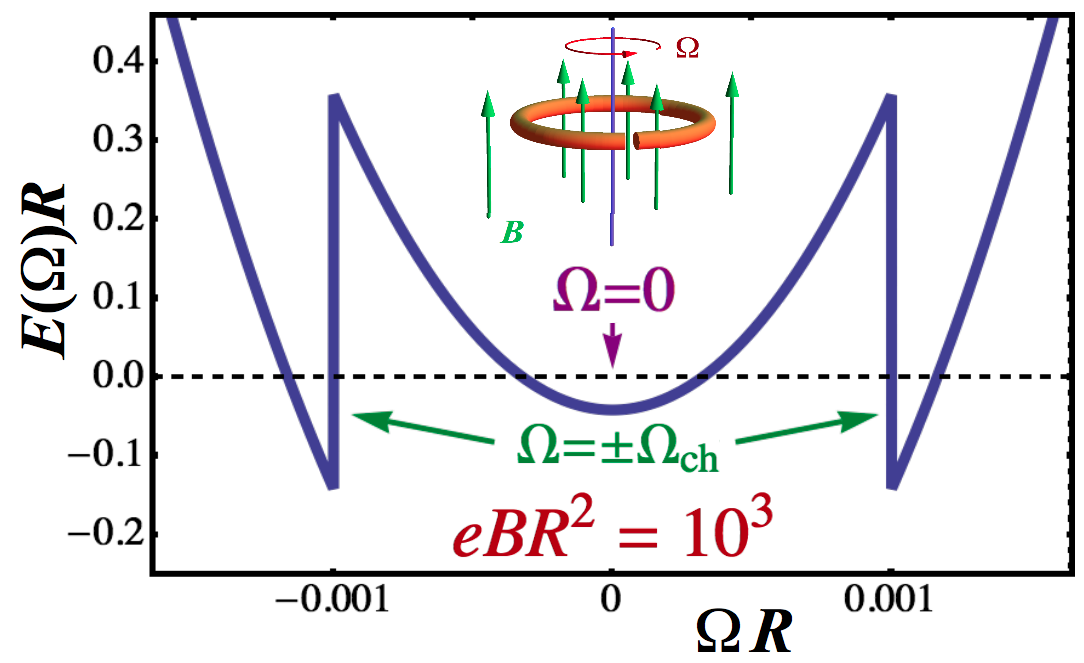} \\[1mm]
\end{tabular}
\end{center}
\vskip -5mm
\caption{{\underline{Left}}: The device which demonstrates that  zero--point vacuum fluctuations have a negative moment of inertia. A free massless scalar field may propagate along the infinitesimally thin ring, while the cut of the ring imposes a Dirichlet boundary condition on this field (from Ref.~\cite{ref:II}). 
{\underline{Right}}: in the magnetic field background $B$ the total (classical plus zero-point) rotational energy of the ring $E = E(\Omega)$ becomes a non-monotonic discontinuous function of the angular frequency~$\Omega$. In this example the minimum of the rotational energy [with $E(\Omega_\ch) < E(0)$] is reached at the nonzero characteristic frequency $\Omega_\ch \neq 0$, Eq.~\eq{eq:Omega:ch}; see the discussion in the text.}
\label{fig:setup}
\end{figure}

In order to illustrate our idea in general, and the rotational vacuum effect in particular, we consider a very simple device; see Fig.~\ref{fig:setup}(left). This toy device contains:
\begin{enumerate}
\item[A.] Classical subsystem: 
 \begin{itemize} 
  \item We consider an infinitely thin (but, generally, massive) ring of the radius $R$. 
  \item The ring may rotate about its natural symmetry axis. 
  \item The ring has a cut.
 \end{itemize}  
 \item[B.] Quantum subsystem:
\begin{itemize} 
 \item The ring is made of a material which possesses massless (scalar) excitations.
 \item The excitations are free (they do not interact with each other).
 \item The excitations may propagate along the ring.
 \item The excitations cannot propagate through the cut of the ring.
\end{itemize}
\item[C.] The classical and quantum subsystems are coupled to each other thanks to the cut.
\end{enumerate}

For the sake of simplicity, we assume that the cut of the ring is infinitesimally thin while the tunnelling of excitations from one side of the cut to another side is absent.

Our next logical steps are as follows:

\begin{enumerate}

\item It is easy to show that the zero-point (vacuum) fluctuations of the excitations in the ring have a negative energy (called, in general, the Casimir energy~\cite{ref:Casimir}). 

\item It turns out that zero-point fluctuations do indeed have a negative moment of inertia (``the rotational vacuum effect'') as was rigorously demonstrated in Refs.~\cite{ref:I,ref:II} and confirmed in Ref.~\cite{Schaden:2012dp}. 

\item The negative moment of inertia implies that the rotational energy of the zero-point vacuum fluctuations decreases as the angular frequency $\Omega$ of the device increases. 

\item However, the rotational vacuum effect -- in its simplest formulation -- is a very small effect~\cite{ref:I}: the (negative) zero-point moment of inertia constitutes a small fraction of the (positive) classical moment of inertia of the device itself.  

\item  Nevertheless, the rotational vacuum effect may be enhanced if
\begin{enumerate}
\item the scalar excitations in the ring are electrically charged; and 
\item the ring is pierced by an external magnetic flux. 
\end{enumerate}

\item Due to the enhancement, the negative zero--point rotational energy may overcome the positive classical rotational energy for certain angular frequencies. Then, the lowest energy state should correspond to a permanently rotating device~\cite{ref:II}.

\end{enumerate}

Below we will briefly discuss the rotational vacuum effect as well as the possible existence of the perpetuum mobile of the fourth kind following Refs.~\cite{ref:I,ref:II}.

\section{The rotational vacuum effect for a neutral scalar field~\cite{ref:I}}

Consider a neutral massless scalar field $\phi = \phi(t,\varphi)$ which is defined on a circle (infinitesimally thin ring) of radius $R$; see Fig.~\ref{fig:setup}(left). The corresponding Lagrangian~is:
\beqn
\cL & = & \frac{1}{2} \partial_{\mu} \phi \, \partial^{\mu} \phi \equiv  \frac{1}{2} \left[\left(\frac{\partial \phi}{\partial t}\right)^{2} - \frac{1}{R^{2}} \left(\frac{\partial \phi}{\partial \varphi}\right)^{2} \right] \,.
\label{eq:circle:L} 
\eeqn
The scalar field is a periodic function $[\phi(t,\varphi + 2 \pi) \equiv \phi(t,\varphi)]$ of the angular coordinate $\varphi \in [0, 2\pi)$. The cut of the ring -- see Fig.~\ref{fig:setup}(left) -- imposes the simplest, Dirichlet boundary condition on the neutral field~$\phi$ (we call the cut as the ``Dirichlet cut''). 

First, let us consider a static, non-rotating circle. The boundary condition at the Dirichlet cut is as follows (the points $\varphi = 0$ and $\varphi = 2 \pi$ are identified):
\beqn
\phi (t,\varphi)\bl_{\varphi = 0} \equiv \phi (t,\varphi)\bl_{\varphi = 2 \pi}  = 0\,.
\label{eq:cirle:D:0}
\eeqn  

It is pretty straightforward to evaluate the zero-point (Casimir) energy associated with the vacuum fluctuations of the scalar field $\phi$.  The local energy density,
\beqn
{\bar {\mathcal E}}(R) = \frac{1}{2 \pi R} \sum_{{\mathrm{modes}}\, m} \varepsilon_{m}(R)\,,
\label{eq:calE:R}
\eeqn 
is given by the infinite sum over energies $\varepsilon_{m}(R)$ of all individual fluctuation modes $m$.  This sum is a divergent quantity both for the circle with a finite radius $R$ and in a free space ($R \to \infty$). However, the difference between these energy densities, ${\bar {\mathcal E}}^{\mathrm{phys}}(R) = {\bar {\mathcal E}}(R) - {\bar {\mathcal E}}(\infty)$, is a finite, experimentally measurable observable which is often called the Casimir energy density~\cite{ref:Casimir}. The total energy of the quantum fluctuations is as follows:
\beqn
E^{\mathrm{phys}}(R) = 2 \pi R\,  {\bar {\mathcal E}}^{\mathrm{phys}}(R) \equiv 2 \pi R  \left[{\bar {\mathcal E}}(R) - {\bar {\mathcal E}}(\infty) \right]\,,
\label{eq:Casimir:energy}
\eeqn

The local energy density of the quantum fluctuations ${\mathcal E}$ is given by the vacuum expectation value of a time-time component of the stress--energy tensor, $T^{\mu\nu}$:
\beqn
{\mathcal E}(t,\varphi) \equiv \left\langle T^{00} (t,\varphi) \right\rangle = \lim_{t' \to t} \lim_{\varphi' \to \varphi} \frac{1}{2 i}
\left(\frac{\partial}{\partial t} \frac{\partial}{\partial t'} + \frac{1}{R^{2}}  \frac{\partial}{\partial \varphi} \frac{\partial}{\partial \varphi'}\right) G(t,t';\varphi,\varphi')\,, \qquad
\label{eq:E:T00}
\label{eq:T00}
\eeqn
where we have expressed the expectation value of the stress--energy tensor $T^{\mu\nu}$,
\beqn
\left\langle T^{\mu\nu} (x) \right\rangle = \left( \partial^{\mu}\partial'^{\nu} - \frac{1}{2} g^{\mu\nu} \partial^{\lambda} \partial'_{\lambda}\right) \frac{1}{i} G(x,x') \bl_{x \to x'}\,,
\label{eq:Tmunu}
\eeqn
via a Feynman--type Green function~\cite{ref:Milton:Book}: $G(x,x') = i \left\langle {\mathrm T} \phi (x) \phi (x') \right\rangle$. Here the symbol ``$\mathrm{T}$'' stands for the time--ordering operator and $x = (t,\varphi)$.

The Green function $G(t,t';\varphi,\varphi')$ is defined by the following equation:
\beqn
\left(\frac{\partial^{2}}{\partial t^{2}} {-} \frac{1}{R^{2}} \frac{\partial^{2}}{\partial \varphi^{2}}\right) G(t,\varphi;t',\varphi') {=} \frac{1}{R} \delta(\varphi - \varphi') \delta(t - t') \,.
\label{eq:Green:function}
\eeqn
The Green's function $G$ should vanish at the position of the Dirichlet cut, $\varphi, \varphi'  = 0,2\pi$.

The Green function $G(t,t';\varphi,\varphi')$ can be expressed via the eigenvalues $\lambda_{\omega,m}$ and eigenfunctions $\phi_{\omega,m}$ of the second--order differential operator in the right hand side of Eq.~\eq{eq:Green:function}:
\beqn
\left(\frac{\partial^{2}}{\partial t^{2}} -  \frac{1}{R^{2}} \frac{\partial^{2}}{\partial \varphi^{2}}\right) \phi_{\omega,m}(t,\varphi) =  \lambda_{\omega,m} \, \phi_{\omega,m}(t,\varphi)\,, \qquad
\label{eq:phi:solutions}
\eeqn
where $m = 1,2,3,\dots$ and $\omega \in {\mathbb R}$, and the orthonormal and complete eigensystem is
\beqn
\qquad
\lambda_{\omega,m} = \frac{m^{2}}{4 R^{2}} - \omega^{2}\,,
\qquad 
\phi_{\omega,m}(t,\varphi) = e^{ - i \omega t} \, \chi_{m} (\varphi)\,, 
\qquad 
\label{eq:phi:omega}
\chi_{m} (\varphi)  =  \frac{1}{\sqrt{\pi R}} \sin \frac{m \varphi}{2}\,.
\label{eq:chi:static} 
\label{eq:eigenvalues}
\eeqn
For the sake of convenience, the real-valued eigenfunctions of Eq.~\eq{eq:phi:solutions} with the same eigenvalues are combined into one complex function $\phi \equiv \phi^{(C)}_{\omega,m} - i \phi^{{(S)}}_{\omega,m}$.

The Green's function is given by the following formula,
\beqn
\qquad
G(t,t';\varphi,\varphi') = \int\limits_{-\infty}^{+\infty} \frac{d \omega}{2 \pi}  \sum_{m=1}^{\infty}
\frac{\phi_{\omega,m}(t,\varphi) \phi^{\dagger}_{\omega,m}(t',\varphi')}{\lambda_{\omega,m} - i \epsilon} 
= \frac{i}{\pi} {\mathcal{G}}\left(\frac{\varphi}{2},\frac{\varphi'}{2}, \frac{| (t - t')|}{2 R} \right)\,,
\label{eq:G:explicit} \label{eq:G:general} \label{eq:explicit:G:1}
\eeqn
where function ${\mathcal{G}}$ is defined as follows:
\beqn
\label{eq:G:cal} 
{\mathcal{G}}(x,y,z) & = & \sum_{m=1}^{\infty} \frac{\sin(m x) \sin(m y) }{m} e^{- i m z}  = \frac{1}{4} \ln \frac{\left[1 {-} e^{i (x + y - z)}\right] 
\left[ 1 {-} e^{- i (x + y + z)}\right]}{\left[ 1 {-} e^{i (x - y - z)}\right] \left[ 1 {-} e^{i (- x + y - z) }\right]}\,.
\eeqn
The infinitesimal term $i \epsilon$ ensures that the Green function~\eq{eq:G:general} is of the Feynman type~\cite{ref:Milton:Book}.

Using the explicit representation of the Green's function~\eq{eq:explicit:G:1} the (non-regularised and divergent) local energy density~\eq{eq:T00} can be computed straightforwardly:
\beqn
\left\langle T^{00}(t,\varphi) \right\rangle  = - \frac{1}{2 \pi} \lim_{t' \to t} \frac{1}{(t' - t)^{2}} - \frac{1}{96 \pi R^{2}}\,.
\label{eq:T0:div}
\eeqn

The energy density~\eq{eq:T0:div} depends on neither  the angular variable~$\varphi$ nor the time variable $t$. As the divergent piece in Eq.~\eq{eq:T0:div} does not depend on the radius of the ring, it is easy to get the regularised expression for the zero--point energy ${E}^{\mathrm{phys}}$ and the mean energy density ${\bar{\mathcal E}}^{\mathrm{phys}}$ of the scalar field at the non-rotating circle with the Dirichlet cut:
\beqn
{E}^{\mathrm{phys}} \equiv 2 \pi R \, {\bar{\mathcal E}}^{\mathrm{phys}} = - \frac{1}{48 R} \,, 
\qquad \quad
{\bar{\mathcal E}}^{\mathrm{phys}} = - \frac{1}{96 \pi R^{2}}\,.
\label{eq:T00:reg:2}
\label{eq:Casimir:total}
\eeqn
The energy vanishes in the limit of the infinitely large circle, $R \to \infty$, as it should be. The result~\eq{eq:Casimir:total} is known as the Casimir energy of the string of the length $l = 2 \pi R$~\cite{ref:Milton:Book,Luscher:1980fr}. 

In order to find the moment of inertia of the zero--point vacuum fluctuations let us repeat the above calculations for the case of the rotating ring, Fig.~\ref{fig:setup}(left), which rotates uniformly about its axis with the angular velocity $\Omega$. Mathematically, the rotation leads to the following time--dependent boundary condition imposed by the Dirichlet cut:
\beqn
\phi (t,\varphi)\bl_{\varphi = {[\Omega t]}_{2\pi}}  = 0\,, 
\label{eq:cirle:D:cond}
\eeqn
where ${[x]}_{2 \pi} = x + 2 \pi n$, $n \in \Z$, $ 0 \leqslant {[x]}_{2 \pi} < 2 \pi$ denotes the modulo operation with the base of $2 \pi$. In the static limit, $\Omega = 0$, the boundary condition~\eq{eq:cirle:D:cond} reduces to Eq.~\eq{eq:cirle:D:0}. The subluminal rotation of the ring ($|\Omega R| < 1$) is always assumed in our calculations.

The Green function in the rotating case should satisfy the general equation~\eq{eq:Green:function} and respect the time-dependent boundary conditions~\eq{eq:cirle:D:cond}. Similarly to Eq.~\eq{eq:G:explicit} the Green function can be expressed via the orthonormal and complete eigensystem of the problem~\eq{eq:Green:function} and \eq{eq:cirle:D:cond} in the laboratory frame: 
\beqn
\qquad
\phi_{m,\omega}(t,\varphi) & = & \sqrt{\frac{1}{\pi R}} \sin \Bigl[\frac{m}{2} {[\varphi - t \Omega]}_{2 \pi} \Bigr] 
\exp\biggl\{ - i \omega \biggl(t - \frac{\Omega R^{2}\, {[\varphi - t \Omega]}_{2 \pi}}{1 - \Omega^2 R^{2}} \biggr)\biggr\}, \quad
\label{eq:phi:m} \\
\lambda_{\omega,m} & = & \frac{1 - \Omega^{2} R^{2}}{4 R^{2}} m^{2} - \frac{\omega^{2}}{1 - \Omega^{2} R^{2}} \,. \qquad 
\label{eq:lambda:omega:m}
\eeqn

The explicit form of the Green function of the rotating ring is as follows~\cite{ref:II}:
\beqn
G_{\Omega}(t,t;\varphi,\varphi') = \frac{i}{\pi} {\mathcal{G}}\left(\frac{\theta(t,\varphi)}{2},\frac{\theta(t',\varphi')}{2}, 
 \frac{\left|(1 - \Omega^{2} R^{2}) (t - t') - \Omega R^{2} \left[\theta(t,\varphi) - \theta(t',\varphi') \right] \right|}{2 R} \right)\!,
\nonumber
\eeqn
where $\theta(t,\varphi) = {[\varphi - \Omega t]}_{2\pi}$ and the function ${\mathcal G}(x,y,z)$ is given in Eq.~\eq{eq:G:cal}. The corresponding zero--point energy density~\eq{eq:T00} is given by the following formula:
\beqn
\left\langle T^{00}(t,\varphi) \right\rangle = - \frac{1}{2 \pi } \lim_{t' \to t} \frac{1}{(t' - t)^{2}} - \frac{1 + \Omega^{2} R^{2}}{96 \pi R^{2}}\,.
\label{eq:T0:div:Omega}
\eeqn
The energy density~\eq{eq:T0:div:Omega} of the rotating ring is space-time independent similarly to the energy density of the static ring~\eq{eq:T0:div}. The divergent piece does not depend on neither the radius of the ring $R$ nor the angular frequency $\Omega$ so that it does not contribute to the physical part of the rotational energy density of zero--point (ZP) fluctuations:
\beqn
{\mathcal E}^{\mathrm{ZP}}_{\Omega} (t,\varphi) \equiv {\left\langle T^{00}(t,\varphi) \right\rangle}^{\mathrm{phys}}  = - \frac{1 + \Omega^{2} R^{2}}{96 \pi R^{2}}\,, 
\qquad\qquad |\Omega R| < 1\,.
\label{eq:T00:rotating}
\eeqn
For a static device, $\Omega = 0$, the energy density of the zero-point vacuum fluctuations~\eq{eq:T00:rotating} reduces to the known result~\eq{eq:T00:reg:2}.

Finally, we arrive to the exact {\it relativistic} expression for the total zero--point energy of the rotating ring with the cut~\cite{ref:I}:
\beqn
E^{\mathrm{ZP}}_{\Omega} & \equiv & R \int\nolimits_{0}^{2 \pi} d \varphi \, {\mathcal E}^{\mathrm{ZP}}_{\Omega} (t,\varphi) = - \frac{1 + R^{2} \Omega^{2}}{48 R}\,, 
\qquad\qquad |\Omega R| < 1\,. 
\label{eq:ZP:B0}
\eeqn
Curiously, the $\Omega$--dependent part of the rotational energy of the relativistic zero-point fluctuations~\eq{eq:ZP:B0} has a very simple form resembling the nonrelativistic classical expression~\eq{eq:E:class}. Result~\eq{eq:ZP:B0} was confirmed in Ref.~\cite{Schaden:2012dp} by using a different method.

The moment of inertia of the zero-point vacuum fluctuations is a negative quantity:
\beqn
I^{\mathrm{ZP}} \equiv \frac{\partial^{2}}{\partial \Omega^{2}} E^{\mathrm{ZP}}_{\Omega} = - \frac{\hbar R}{24 c}\,,
\label{eq:I:ZP:no:enhancement}
\eeqn
where we have restored the Planck constant $\hbar$ and the speed of light $c$. The energy of the zero-point vacuum fluctuations decreases, as the angular frequency increases. Thus, the classical positivity condition~\eq{eq:I:cl} is not valid for the zero-point vacuum fluctuations~\eq{eq:I:ZP:no:enhancement}.

Notice that despite the fact that $I^{\mathrm{ZP}} < 0$, a rotation of an {\it isolated} device cannot self--accelerate because of the conservation of the angular momentum (in order to change its angular momentum, the device should exchange the angular momentum with environment by, say, emitting or absorbing a photon).

Equation~\eq{eq:I:ZP:no:enhancement} provides us with the lowest bound~\cite{Schaden:2012dp} on the moment of inertia for the ring-type devices [Fig.~\ref{fig:setup}(left)] driven by the neutral scalar vacuum. The zero--point moment of inertia~\eq{eq:I:ZP:no:enhancement} is an very small quantity, corresponding to a fractional (in terms of $\hbar$) angular momentum~\cite{Schaden:2012dp}. Thus, the positive moment of inertia of a real massive device~\eq{eq:I:cl:m} is expected to be always greater than the negative moment of inertia of the zero-point fluctuations~\eq{eq:I:ZP:no:enhancement}.

\section{Enhancement of the rotational vacuum effect by magnetic field~\cite{ref:II}}

The negative moment of inertia can be drastically enhanced by an external magnetic field $B$ if the massless particles are electrically charged~\cite{ref:II}. The corresponding zero-point energy $E^{\mathrm{ZP}}_{\Omega,B}$ and the zero-point angular momentum ${L}^{\mathrm{ZP}}_{\Omega,B}$ of a free charged field:
\beqn
\qquad
E^{\mathrm{ZP}}_{\Omega,B} & \equiv & R \int\nolimits_{0}^{2 \pi} d \varphi\,  \left\langle T^{00}(t,\varphi) \right\rangle
= - \Bigl[1 + 6  M_{\Omega,B} (M_{\Omega,B} + 1)\Bigr]  \frac{1 + \Omega^{2} R^{2}}{24 R} \,, 
\label{eq:E:rotating:B}
\\
{L}^{\mathrm{ZP}}_{\Omega,B}  & \equiv & 
R \int\nolimits_{0}^{2 \pi} d \varphi\, \left[ x^{2} \left\langle T^{10} \right\rangle -  x^{1} \left\langle T^{20}  \right\rangle \right]
= - \Bigl[1 + 6  M_{\Omega,B} (M_{\Omega,B} + 1)\Bigr]  \frac{\Omega R}{12} \,,
\label{eq:Tphi0:rotating:B}
\eeqn
are related to each other by a ``classical'' relation:
\beqn
L^{\mathrm{ZP}}_{\Omega,B} = \frac{\partial E^{\mathrm{ZP}}_{\Omega,B}}{\partial \Omega} \bl_{\frac{e B \Omega R^{3}}{1 - \Omega^{2} R^{2}} \notin \Z}\,, 
\label{eq:L:ZP}
\qquad  \ \mbox{where} \qquad
M_{\Omega,B} = \left\lfloor \frac{e B \Omega R^{3}}{1 - \Omega^{2} R^{2}} \right\rfloor \in \Z \,,
\eeqn
provided the angular frequency $\Omega$ and the magnetic field $B$ do not correspond to the discontinuities in the integer-valued quantity $M_{\Omega,B}$ (here $\lfloor x \rfloor$ is a largest integer which is smaller than $x$). In the limit of vanishing magnetic field the charged-field result \eq{eq:E:rotating:B} turns into the neutral-field expression~\eq{eq:ZP:B0} multiplied by a factor of two because the charged (complex) field has two degrees of freedom. In the nonrelativistic limit, $\Omega R \ll c$,  the first discontinuity of the zero-point energy~\eq{eq:E:rotating:B} happens at the characteristic frequency,
\beqn
\Omega_{\ch}(B) = \frac{\hbar c}{e B R^{3}}\,.
\label{eq:Omega:ch:hbar} \label{eq:Omega:ch}
\eeqn
A global minimum of the total (classical plus zero-point) rotational energy is reached, generally, at a nonzero optimal frequency, $\Omega = n\, \Omega_\ch$ with $n \in\Z$ (up to small relativistic corrections). The illustrative example of the total rotational energy, shown in Fig.~\ref{fig:setup}(right), corresponds to $\Omega = \pm \Omega_\ch \neq 0$; see Ref.~\cite{ref:II} for further details.

The magnetic field strongly enhances the zero-point energy~\eq{eq:E:rotating:B} because the integer number $M_{\Omega,B}$ may become very large. In Ref.~\cite{ref:II} a possible application of this approach to torus-shaped carbon nanotubes was discussed and it was shown that the enhancement factor may reach astronomically high values ($10^{15}$ and higher). Notice, however, that in computing of Eq.~\eq{eq:E:rotating:B} we have assumed that the charged particles are free, so that they may interact with the external magnetic field while the photon exchange between them is absent. The importance of the photon-mediated interaction is unclear for a moment.

\begin{figure}[!thb]
\begin{center}
\includegraphics[scale=0.08,angle=-90,clip=false]{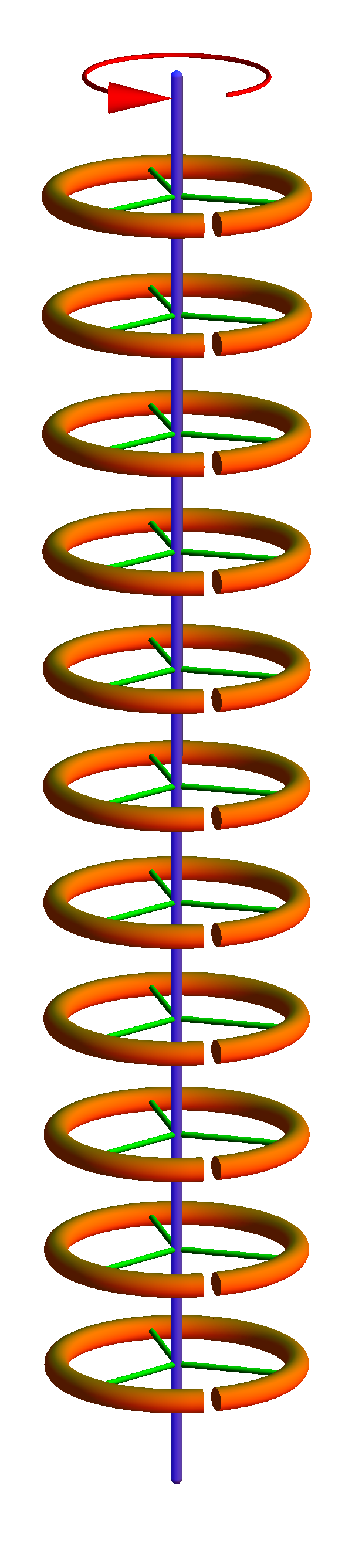}
\end{center}
\vskip -5mm
\caption{A macroscopic permanently rotating device is visually similar to the very first proposal of a metamaterial~\cite{ref:split:ring} made of the C--shaped split-ring resonators (from Ref.~\cite{ref:II}).}
\label{fig:multidevice}
\end{figure}

Finally, we notice that the device may be made macroscopically large by assembling the open rings along their common axis in a metamaterial-like~\cite{ref:split:ring} fashion; see Fig.~\ref{fig:multidevice}.

\section{Perpetuum mobile of the fourth kind and thermodynamics}

The perpetuum mobile of the fourth kind does not violate the laws of thermodynamics (see Ref.~\cite{ref:II} for a more detailed discussion):

\begin{itemize}

\item[I.] The very existence of this device is obviously consistent with the first law of thermodynamics because no work is produced by the object which rotates in its lowest--energy ground state (otherwise such a device would be the perpetuum mobile of the {\underline{first kind}} which is obviously forbidden by the energy conservation law). 

\item[II.] Because of the same reason, our device does not spontaneously convert thermal energy into mechanical work so that the second law of thermodynamics is not violated as well  (otherwise such a device would be the perpetuum mobile of the {\underline{second kind}} which is forbidden too). 

\item[III.] And, because of the very same reason, this device cannot serve as a perpetual energy storage (thus our device is not the perpetuum mobile of the {\underline{third kind}}). 

\item[IV.] Our device is the perpetuum mobile of the {\underline{fourth kind}}: it does not produce work and it cannot serve as a perpetual energy storage. But it rotates perpetually.

\end{itemize}

Can a permanently rotating macroscopic body be in thermodynamic equilibrium with the external environment?\footnote{We thank G.~E.~Volovik for rising this question.} In the equilibrium, the angular velocity $\boldsymbol \Omega$, angular momentum $\boldsymbol L$, energy $E$ and entropy $S$ of the body are related to each other as follows~\cite{ref:LL5}:
\beqn
{\boldsymbol{\Omega}} = {\left(\frac{\partial E}{\partial {\boldsymbol L}}\right)}_{S}\,.
\label{eq:Omega:E}
\eeqn
If the energy of the system is a smooth function of the angular momentum, $E = E({\boldsymbol L})$, then in the equilibrium the angular velocity of the body should always be zero, ${\boldsymbol \Omega} = 0$ (this statement works, obviously, at any temperature including the absolute zero). 

Thus, we arrive to the important conclusion~\cite{ref:II}: 
\vskip 3mm
\begin{center}
{\it ~\phantom{phant}the rotational energy of a permanently rotating device must be \newline a discontinuous single-valued function of its angular momentum}. 
\end{center}
\vskip 3mm
Remarkably, this property is always satisfied for our device due to the presence of the discontinuities in the zero-point contributions to the energy~\eq{eq:E:rotating:B} and to the angular momentum~\eq{eq:Tphi0:rotating:B}; see Fig.~\ref{fig:setup}(right) for illustration. Moreover, in Ref.~\cite{ref:II} we have explicitly demonstrated that, for our device, the thermodynamic relation~\eq{eq:Omega:E} does indeed give rise to nonzero angular velocity at the equilibrium, ${\boldsymbol \Omega} \neq 0$. Thus, the existence of the perpetuum mobile of the fourth kind is consistent with the laws of thermodynamics.

\clearpage

\section{Conclusions}

We have demonstrated that zero--point vacuum fluctuations may possess a negative moment of inertia. This fact means that the rotational energy of the zero--point fluctuations decreases with the increase of the angular frequency (``the rotational vacuum effect''~\cite{ref:I}).  The presence of a magnetic field background may drastically enhance the negative moment of inertia of zero--point fluctuations so that the negative rotational energy of the zero--point fluctuations may compensate the positive classical rotational  energy of the device itself~\cite{ref:II}. In this case the device -- which does not produce any work -- becomes a perpetuum mobile of the fourth kind driven by the zero--point fluctuations. 

\acknowledgments 

The author is grateful to M.~Asorey, E.~Elizalde, K.~Kirsten, K.~Milton, M.~Plyushchay, M.~Schaden, D.~Vassilevich and G.~E.~Volovik for interesting discussions. The work was supported by Grant No. ANR-10-JCJC-0408 HYPERMAG (France).

\end{document}